# The Dichotomy of Distributed and Centralized Control: METRO-HAUL, when control planes collide for 5G networks

D. King [(1)] *, A. Farrel [(1)], Emiko Nishida-King [(1)], R. Casellas [(2)], L. Velasco [(3)], R. Nejabati [(4)], A. Lord [(5)]

1. Old Dog Consulting (ODC), Wales, United Kingdom
2. Centre Tecnològic de Telecomunicacions de Catalunya (CTTC/CERCA), Castelldefels, Spain
3. Optical Communications Group (GCO), Universitat Politècnica de Catalunya (UPC), Barcelona, Spain.
4. University of Bristol, Bristol, United Kingdom
5. British Telecom, United Kingdom

*Corresponding author: daniel@olddog.co.uk

*Abstract*—Automating the provisioning of 5G services, deployed over a heterogeneous infrastructure (in terms of domains, technologies, and management platforms), remains a complex task, yet driven by the constant need to provide end-to-end connections at network slices at reducing costs and service deployment time. At the same time, such services are increasingly conceived around interconnected functions and require allocation of computing, storage, and networking resources.

The METRO-HAUL 5G research initiative acknowledges the need for automation and strives to develop an orchestration platform for services and resources that extends, integrates, and builds on top of existing approaches, macroscopically adopting Transport Software Defined Networking principles, and leveraging the programmability and open control of Transport SDN.

## 1 INTRODUCTION AND MOTIVATION

All agree that 5G and Internet of Everything (IoE) services will have a significant impact on traffic, including the types, volume, and dynamicity of traffic, while at unprecedented transmission rates [1]. To facilitate these emerging traffic requirements, the optical transport network should become more responsive to traffic changes as well as operating more efficiently. Key enablers include Software Defined Networking (SDN) and Network Functions Virtualisation (NFV): combined they promise the increased network flexibility and automation needed to deliver Transport SDN for 5G and beyond [2].

The METRO-HAUL project is a European Commission funded project that involves the design and development of a novel, spectrally efficient, and adaptive network solution using dynamic elastic optical networking, including both transparent and flexible optical switching and adaptive transmission. METRO-HAUL will address the granularity mismatch between the 5G access and the optical metro domain and achieve dynamic optical bandwidth assignment and allocation. This will provide metro support for the increased volume of services with reduced cost and energy consumption.

To support the required dynamicity and flexibility highlighted previously, the METRO-HAUL architecture will need to integrate a wide range of Transport SDN technologies. These will be controlled using automation schemes and programmability features that will enable disaggregation and virtualization concepts, the coordination of which will be supported by a control plane designed for the purpose. This new control plane will dynamically adapt to the needs of specific services, optimally exploiting the data plane through the use of relevant data monitoring and analysis schemes. The control plane will be also responsible for the provisioning of 5G and IoE industry services and ensure the required end-to-end QoS and QoE levels for each service. Therefore, the METRO-HAUL control plane will have to leverage well-established distributed control and signaling methods, whilst utilising emerging SDN and NFV paradigms, somehow unifying the whole to exploit the benefits of a unified system.

The following sub-sections outline the key 5G service control plane requirements for enabling the Transport SDN infrastructure, as research challenges, developed by the METRO-HAUL team.

### 1.1 A Need for Automation

The provisioning and operation of optical transport connections, both accurately and via an automated interface, including path feasibility evaluations, are critical requirements for METRO-HAUL. Increasingly it is expected that programming of the network will need to be performed via network APIs, using model-based technologies. The IT domain has used resource modeling methods, but their application to network infrastructure operations are a very recent development. Their use for optical networking, and specifically the 5G requirements, needs to be investigated.

*1.2    Efficient Service Placement*

These connections often require the evaluation and assessment of the quality of the optical channels available and depend on several aspects such as link lengths, type of fibres, number and type of Network Elements (NE) traversed, bit rate, modulation format, wavelength channel, economic cost, and physical reliability. Path selection often involves complex and time-consuming computations. Strategies for online path feasibility evaluation rely on a compromise between accuracy, computation time, and the amount of information required.

Two approaches are typically used for path selection:

- Firstly, a priori assessment by having the feasible paths and constraints computed in advance.
- Secondly, real-time (relatively speaking) computation in response to changing network conditions.

The first approach is performed offline and allows global optimisation factors to be applied:as network complexity increases, computing power may need to be increased exponentially to evaluate the search space being considered. The second approach has the main disadvantage that the speed of calculation and the response processing may delay the setup of the service. Simplifications may be applied that reduce the complexity, but the accuracy and optimality of the path computation may be negatively affected.

In both scenarios, the amount of information that needs to be imported and processed can become very large (e.g., in large networks, with a high number of possible paths, modulation formats, bit rates, etc.), which can hamper the scalability of either approach [3].

*1.3    Network Slicing*

Network slicing in 5G systems defines logical, self-contained networks that consist of a mixture of shared and dedicated resource instances [4]. This may include radio spectrum, network equipment, compute resources, and storage resources. We widen this definition to include network abstraction, a technique that can be applied to a network domain to select network resources by policy to obtain a view of potential connectivity [5].

The METRO-HAUL network slicing requirement is an approach to network operations that builds on the concept of network abstraction and programmability. It will use SDN and NFV to create multiple logical (virtual) networks, each tailored for a set of 5G services that share similar requirements and operate on top of a common network. If achieved, this networking and service flexibility would represent a radical change, beyond network sharing, enabling tailored 5G services to be delivered to third parties and vertical market players.

No single, concise, definition of "network slice" exists for 5G networks that utilise an enabling optical infrastructure. There are several descriptions, across Standards Development Organisations, that may have elements of applicability to METRO-HAUL. However, these are all biased towards a particular technology domain. It was critically important that METRO-HAUL did not limit itself to any specific definition of application or technology. Thus, the METRO-HAUL network slice is born:

> A "Network Slice" represents an agreement between a User and a Service Provider to deliver network resources according to a specific service level agreement. In this context, a "User" is an application, client network, or customer of a Service Provider. And a "Service Provider" is a network operator that controls a server network or a collection of server networks.
>
> "Network resources" are any features or functions that can be delivered by a server network. This includes connectivity, compute resources, storage, and content delivery. A "service level agreement" describes multiple aspects of the agreement between the user and the service provider:
>
> - the quality with which the features and functions are to be delivered including measures of bandwidth, latency, and jitter;
>
> - the types of service (such as the network service functions or billing to be executed);
>
> - the location, nature, and quantities of services (such as the amount and location of computing resources and the accelerators require).

A network slice does not necessarily represent dedicated resources in the server network but does constitute a commitment by the service provider to provide a specific level of service. Thus, a network slice could be realised as virtualisation of (consider virtual private wires and VPNs), or as partitioning and dedication of server network resources.

A network slice can further be a detailed description of a complex service that will meet the needs of a set of applications. Such a service may be requested dynamically (that is, instantiated when an application requires it and released when the application no longer needs it), and modified as the needs of the application change.

*1.3.1 Requirements for Network Slicing*

It is expected that METRO-HAUL network slicing capabilities will have to provide:

*Resource Slicing* - For network slicing, it is important to consider both infrastructure resources and service functions as described in Figure 1: "METRO-HAUL Disaggregation Architecture". This allows a flexible approach to deliver a range of 5G services both by partitioning (slicing) the available network resources to present them for use by an application or consumer, and also by providing instances of services and network functions at the right locations and in the correct chaining logic, with access to the necessary hardware, including specific compute and storage resources. Mapping of resources to slices may be 1-to-1, or resources might be shared among multiple slices.

*Network and Function Virtualization* - Virtualization is the abstraction of resources where the abstraction is made available for use by an operations entity, for example, by the Network Management Station (NMS) of a high-layer network. The resources to be virtualized can be physical or already virtualized, supporting a recursive pattern within different abstraction layers. Therefore, virtualization will be critical for network slicing as it enables effective resource sharing between network slices.

Just as server virtualization makes virtual machines (VMs) independent of the underlying physical hardware, network virtualization enables the creation of multiple, isolated virtual networks that are completely decoupled from the underlying physical network and can safely run on top of it.

*Resource Isolation* - Isolation of data and traffic is a major requirement that must be satisfied for certain applications to operate in concurrent network slices on a common shared underlying infrastructure. Therefore, isolation must be understood in terms of:

- Performance: Each slice is defined to meet specific service requirements, usually expressed in the form of Key Performance Indicators (KPIs). Performance isolation requires that service delivery on one network slice is not adversely impacted by congestion and performance levels of other slices;

- Security: Attacks or faults occurring in one slice must not have an impact on other slices. Service flows are not only isolated on the network edge, but traffic from multiple customers is not mixed in the core of the network;

- Management: Each slice must be independently viewed, utilised, and managed as a separate network;

- o Automation: To minimize dependency on human administrators, the concept of autonomic networking entails closing control loops aiming at providing self-management capabilities, as well as deploying data analytics systems to provide at the analysis of network performance data and discovering usable knowledge (Knowledge Discovery from Data, KDD)

Orchestration is the overriding control method for network slicing. We may define orchestration as combining and coordinating multiple control methods to provide an operational mechanism that can deliver services and can control underlying resources. In a network slicing environment, an orchestrator is needed to coordinate disparate processes and to provide resources for creating, managing, and deploying the end-to-end service. Two scenarios are outlined below where orchestration would be required:

- Multi-Domain Orchestration: Managing connectivity and setup of the transport service across multiple administrative domains;

- End-to-end Orchestration: Combining resources for an "end-to-end service" (e.g., transport connectivity with firewalling and guaranteed bandwidth and minimum delay for premium 5G users or applications spanning multiple domains).

One method of network abstraction being investigated by the METRO-HAUL team is the IETF management and control framework and function proposal, Abstraction, and Control of Traffic Engineered Networks (ACTN) [6].

How the ACTN framework may be applied and extended to meet the goals of METRO-HAUL, providing network and service abstraction and coordination of resources across multiple domains and layers, will take significant investigation, development, and deployment consideration [5].

## 2  ARCHITECTURE

Initial investigations for the METRO-HAUL architecture required consideration of overall network responsibility, node architecture, Optical Infrastructure Element (OIE) capabilities, and targeted 5G services. To facilitate this, a recently proposed concept of "resource disaggregation" will be pivotal. Resource disaggregation relies on physically decoupling components and functions, hosting them at remote locations, instead of coupling all components into a single platform. Thus, disaggregation would enable independence across functions and technologies, providing granular control of resources and how they are programmed and operated.

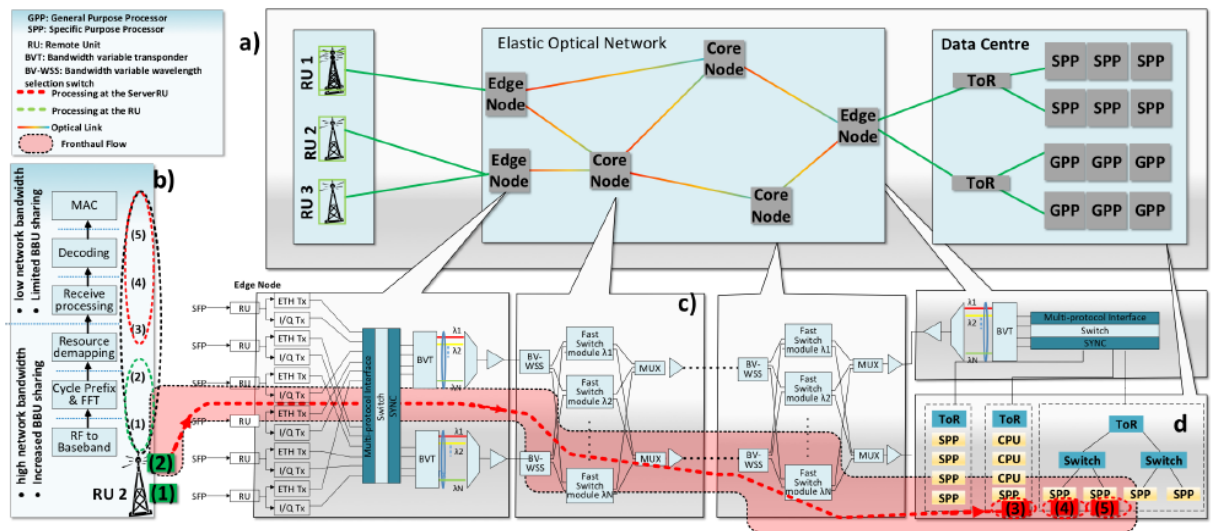

Figure 1: METRO-HAUL Disaggregation Architecture

METRO-HAUL proposes evolving from the traditional Radio Access Network (RAN) architecture to the "Disaggregated RAN" (D-RAN) approach (shown in Figure 1: "METRO-HAUL Disaggregation Architecture"). This would utilise the concepts of 5G and optical hardware and software component disaggregation, enabling METRO-HAUL infrastructure components via a common pool of resources that can be independently selected and allocated [7] based on network services and demand. This would also meet the objective of increased flexibility, scalability, and sustainability relevant to 5G services.

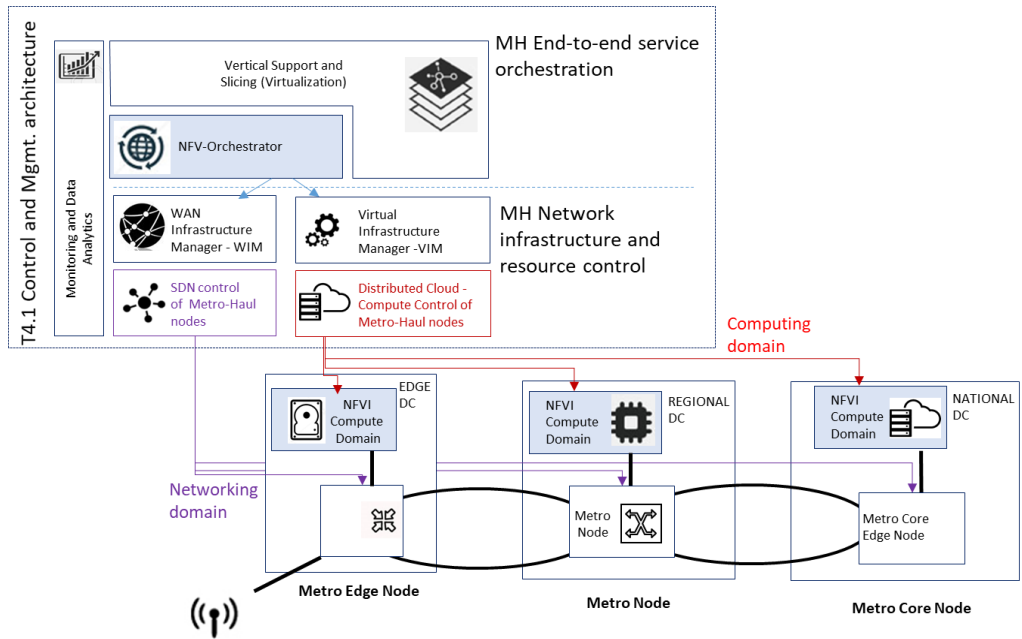

Figure 2: METRO-HAUL Control and Management

Several METRO-HAUL architectural approaches for control and orchestration, including the proposal in Figure 2: "METRO-HAUL Control and Management", will need to be investigated and their suitability analysed. Furthermore, three main principles will also need to be considered: *i) centralized*, *ii) distributed*, and *iii) hierarchical*. The strengths and weaknesses of these principles [7] are described below in Table 1: Centralized, Distributed and Hierarchical Architectures, as are several control functions required for METRO-HAUL.

| Architecture | Features | Strengths | Weaknesses |
| --- | --- | --- | --- |
| **Centralized** | <ul><li>Global view of network resources</li><li>Vendor and technology data plane agnostic</li></ul> | <ul><li>No need for node control plane intelligence or state</li><li>New southbound APIs can be supported directly from the centralized controller</li></ul> | <ul><li>May not reflect rapid state changes in distributed network notes</li><li>Service setup scalability in large networks</li><li>Single point of failure</li></ul> |
| **Distributed** | <ul><li>Highly-available by design as no single-point-of-failure</li><li>Policies can be applied locally at the node level</li></ul> | <ul><li>Significantly better scalability</li><li>Easier to implement protection mechanisms at local node interfaces</li></ul> | <ul><li>No global network resource view</li><li>Computational resources for control plane actions required locally</li></ul> |
| **Hierarchical** | <ul><li>Overall global abstracted view of network resources</li><li>Capable of integrating new lower-layer technologies</li></ul> | <ul><li>Scalable</li><li>Delegates technology specific control to child controllers.</li></ul> | <ul><li>The top-level controller may still represent a single point of failure</li><li>System complexity is increased</li></ul> |

Table 1: Centralized, Distributed, and Hierarchical Architectures.

## 3   OPTICAL CONTROL PLANE

Traditional optical transport networks are proprietary, integrated and closed, where the entire transport network acts as a single vendor managed domain. It can export high-level interfaces and an open North-Bound Interface (NBI), yet the internal details and interfaces are hidden from the operator.

The objective of METRO-HAUL is the disaggregation of optical networks: that refers to a deployment model of optical systems that composes and assembles open and available components, devices, and sub-systems. This disaggregation can be partial or total (down to each of the optical components) and is driven by multiple factors, notably:
- the mismatch between the needs of operators and the ability to deliver adapted solutions by vendors;
- the increase in hardware commoditization;
- the different rate of innovation for different components;
- the promised acceleration on the deployment of services and the consequent reduction in operational and capacity expenses.

Disaggregation imposes a new set of challenges in its control and management. It is clearly a use case for open interfaces that export programmability, and the increase of unified and systematic information and data modelling activities is a crucial step in this regard. However, optical networks are particularly challenging to model due to the lack of agreed-upon hardware models, and this is critical for the development of an interoperable ecosystem around disaggregated hardware.

Recent research [3] has developed control planes to meet the evolving requirements for managing elastic optical infrastructure. Each supports a set of basic functions, including i) element addressing; ii) dynamic resource discovery (e.g., local interfaces and device ports and capabilities); iii) automatic topology and reachability discovery and management (by which a control plane may discover the topology without explicit pre-configuration), iv) path computation, and v) actual service provisioning with recovery (protection and restoration) ensuring efficient resource usage.

In the context of METRO-HAUL, we should underline that 5G access, optical-based infrastructure, disaggregation of resources, network slicing, and enabling end-user control (e.g., User-Network-Interface services) all add significant additional complexity. Therefore an over-arching control philosophy will be required covering all network segments down to the data-centre.

*3.1 Control Plane Design and Control Methods*

The selection and development of a METRO-HAUL control plane must address the requirements highlighted, and yet provide seamless communication, with the functions and responsibilities defined by the architecture selected. A key decision will be the selection of a control plane model based on distributed or centralized architectural principles previously discussed, although a hybrid model (combining centralized and distributed elements) may be feasible.

*3.1.1 Distributed Control Plane*

In a distributed control plane model, each network node has the necessary logic (a control plane entity) to communicate with other network nodes (with logic components). These logic components combine resource discovery, reachability, signaling, and often connection or link management functions.

Each distributed node is responsible for the dissemination of resources under its control (e.g., its own links), so the network view is built in a cooperative way. Once a connection between nodes is required, a service setup is requested. The ingress node is typically responsible for the path computation function based on the topology it has obtained, and for triggering the signaling process by which resources are reserved for the service setup. There is no central authority that coordinates the network operation in a distributed control plane environment.

*3.1.2 Centralized Control Plane*

In a centralized control plane, a controller interacts with the nodes directly. The logic (and topology) remains in the controller, addressing the complexity and cost of distributed control planes. While this architecture simplifies the implementation of the control logic, it has scalability limitations as the size and dynamics of the network increase.

Both models have their strengths and weaknesses: a central control is conceptually simpler, a single point of deployment of policies and business logic, easier to deploy, and requires less state synchronization. It may also present a bottleneck or single point of failure, with latent fault-tolerance issues.

Network functions requiring local knowledge (dynamic restoration, fast rerouting) are harder to achieve in a centralized model, where a distributed model is potentially faster (capable of responding to local knowledge) and more robust and mature, although implementations usually need to conform to a wider set of protocols.

*3.1.3  Which is best, distributed or centralized?*

The control plane (the definition of routing and traffic engineering policy) remains a significant operational task in Transport SDN, and the control of resources via a centralized platform would provide a global network view and efficient use of resources. However, any changes to physical optical network parameters would need to be reflected to the central controller quickly, or it may suffer from scalability problems and compute paths on outdated information.

Distributed control planes adapt quickly to changing conditions so provide high survivability, fast recovery, and can maintain accurate state. However, there is a need to have better configuration management, a clear separation of configuration and operational data for the network slicing objectives as outlined earlier in the document, while enabling high-level constructs more adapted to 5G services and supporting network-wide transactions such as global concurrent optimizations [8].

Therefore, it is not a question of which is best, distributed, or centralized? The question is how we might blend control plane architectures and principles for optimal Transport SDN in 5G networks.

*3.1.4  CAP Theorem Considerations*

We will also need to consider state management across centralized and distributed control plane systems. The CAP theorem [9] views a distributed computing system as composed of multiple computational resources (i.e., CPU, memory, storage) that are connected via a communications network and together perform a task. The theorem identifies three characteristics of distributed systems that are universally desirable:

- *Consistency*, meaning that the system responds identically to a query no matter which node receives the request (or does not respond at all);

- *Availability*, i.e., that the system always responds to a request (although the response may not be consistent or correct);

- *Partition Tolerance*, namely that the system continues to function even when specific nodes or the communications network fail.

A METRO-HAUL centralized controller may act as a consistent global database and specific network mechanisms to ensure new traffic or service requests are handled consistently. If a node cannot reach the controller, the traffic classification will be unavailable until the connection to the controller is restored. Multiple centralized controllers may be deployed, to improve partition tolerance, but at the cost of loss of absolute network resource consistency.

By its nature, a distributed control plane will be dynamic, with any link or service state change being propagated via the distributed communication mechanisms. If we consider convergence after partition of the network, a traditional distributed control-plane operation is usually local and fast (available), while a centralized controller may be slower. Therefore, how we combine a centralized and distributed control plane for overall resource and end-to-end service management in METRO-HAUL environments will need to be considered and overcome.

## 4   CANDIDATE ARCHITECTURE

The proposed METRO-HAUL architecture for Transport SDN will require a top-level overarching Transport SDN control entity which is called the SDN Orchestrator. The SDN Orchestrator addresses over-arching control across multiple heterogeneous domains (both in terms of control and data plane domains). It will be recursive and relies on a hierarchical control structure, which can support both centralized and distributed control planes (via "Child Controllers") when required.

Control of disaggregated optical network components may be performed by the lower centralized SDN controller, as directed by the SDN Orchestrator. Additional lower-layer controllers will provide path computation and traffic engineering functions and may use NETCONF [10] as the configuration protocol and YANG [11] for resource modelling, supporting direct configuration, as outlined in Figure 3: "METRO-HAUL Centralized and Distributed Control Plane Architecture" below.

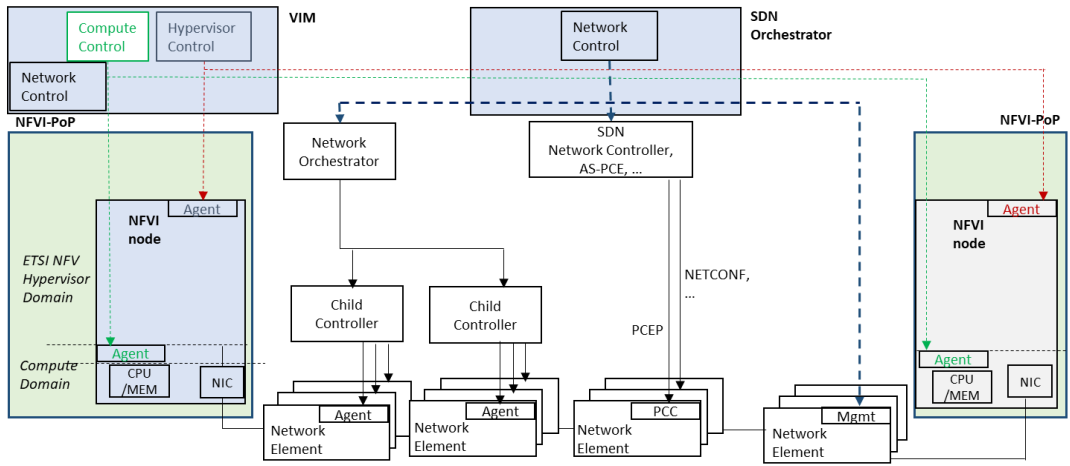

Figure 3: METRO-HAUL Centralized and Distributed Control Plane Architecture.

The Path Computation Element Protocol (PCEP) [12] will be used to request connection setup across distributed control plane environments. By combining both a distributed control and centralized control plane, we can maximise the benefits of both architectures in a hierarchical deployment. The SDN Orchestrator is used for global policy functions, such as: resource optimisation, multi-layer traffic engineering and computing restoration solutions in advance for network failures. The distributed control plane would then be used for local path computations and service recovery in situations that pre-computed restoration schemes are not sufficient. Furthermore, this hierarchical structure is intended to improve scalability performance by reducing the number of NETCONF sessions from the SDN Orchestrator which would be required for management of the METRO-HAUL optical network elements of the (disaggregated) optical infrastructure.

The control elements of the METRO-HAUL SDN Orchestrator and child controllers will utilize key YANG-based resource models. The following YANG model efforts have been identified as candidates: OpenROADM [13] as general reference for ROADM devices (i.e., not including all details); OpenConfig [14] as general reference for XPONDER devices (i.e., not including all details); IETF (including TE-TOPOLOGY [15], WSON [16] and Flexi-Grid [17]), for specific augmented versions of YANG models.

## 5 IMPACT OF METRO-HAUL

The METRO-HAUL project is currently 18-months into its three-year period. Already the project has published numerous papers [18] at key conferences, journals and demonstrated proof-of-concepts for 5G infrastructre control. One notable proof-of-concept demonstration was the automated provisioning of carrier Ethernet over a disaggregated WDM network using a hierarchical SDN control & monitoring platform at ECOC 2018 [19].

## 6 CONCLUSIONS AND FUTURE WORK

The provisioning of 5G services (network connectivity and services involving heterogeneous resources) and network slicing for METRO-HAUL will require automated connection setup to satisfy specific requirements in terms of quality of service, latency, bandwidth, enabling recovery (local protection and pre-computed restoration), across multiple domain and technology layers, using a hierarchical distributed control plane architecture.

While the initial findings on the functional benefits of the METRO-HAUL control plane deployment framework look promising, adopting an approach where both the hierarchical centralized and distributed models [20] can be utilized and exploited will be complex but would ultimately yield the greatest benefits. However, several challenges will stem from stitching heterogeneous environments across multiple technological and administrative domain-levels, spanning multiple resource segments. These challenges include scaling the METRO-HAUL control architecture, addressing the potential system complexity of maintaining state synchronisation between the SDN Orchestrator and Child Controllers, and adapting YANG resource models for control of end-to-end services.

Other technology innovations related to the METRO-HAUL optical control plane include pro-active network monitoring, where it is expected that advances related to data analytics (telemetry) [21] and machine learning will be used to improve control of 5G services managed by the Metro-Haul hierarchical system. A fundamental challenge is the typically static nature, and inflexible design principles of optical networks and connectivity resource control, leading to resource under-utilization, and finally lack of granular resource management and bit-rate flexibility that is required for 5G services. This must also be addressed using novel METRO-HAUL control plane architecture.

## ACKNOWLEDGMENTS

The research leading to this positioning paper has received funding from the European Commission for the H2020-ICT-2016-2 METRO-HAUL project (G.A. 761727) [22].